\documentclass{article}

\usepackage{microtype}
\usepackage{graphicx}
\usepackage{subfigure}
\usepackage{booktabs} 
\usepackage{diagbox}
\usepackage{enumitem}

\usepackage[table]{xcolor}
\usepackage{pgf}          

\usepackage{hyperref}

\usepackage[accepted]{icml2025}

\usepackage{amsmath}
\usepackage{amssymb}
\usepackage{mathtools}
\usepackage{amsthm}
\usepackage{ dsfont }
\usepackage{multirow}

\usepackage[capitalize,noabbrev]{cleveref}

\theoremstyle{plain}

\theoremstyle{definition}

\theoremstyle{remark}

\usepackage[textsize=tiny]{todonotes}

\icmltitlerunning{IMPACT: Iterative Mask-based Parallel Decoding for Text-to-Audio Generation with Diffusion Modeling}

\begin{document}

\twocolumn[
\icmltitle{IMPACT: Iterative Mask-based Parallel Decoding for Text-to-Audio Generation with Diffusion Modeling}



\icmlsetsymbol{equal}{*}
\icmlsetsymbol{amazondisclaimer}{$\dagger$}

\begin{icmlauthorlist}
\icmlauthor{Kuan-Po Huang}{ntu,amazon,amazondisclaimer}
\icmlauthor{Shu-wen Yang}{ntu,amazon,amazondisclaimer}
\icmlauthor{Huy Phan}{amazon}
\icmlauthor{Bo-Ru Lu}{amazon}
\icmlauthor{Byeonggeun Kim}{amazon}
\icmlauthor{Sashank Macha}{amazon}
\icmlauthor{Qingming Tang}{amazon}
\icmlauthor{Shalini Ghosh}{amazon}
\icmlauthor{Hung-yi Lee}{ntu}
\icmlauthor{Chieh-Chi Kao}{amazon}
\icmlauthor{Chao Wang}{amazon}
\end{icmlauthorlist}

\icmlaffiliation{ntu}{National Taiwan University, Taipei, Taiwan}
\icmlaffiliation{amazon}{Amazon AGI, United States}

\icmlcorrespondingauthor{Kuan-Po Huang}{gerber861017@gmail.com}
\icmlcorrespondingauthor{Chieh-Chi Kao}{chiehchi@amazon.com}

\icmlkeywords{Text-to-audio, Diffusion models, Iterative parallel decoding, Mask-based generative modeling}

\vskip 0.3in
]



\printAffiliationsAndNotice{\AmazonDisclaimer} 

\begin{abstract}
Text-to-audio generation synthesizes realistic sounds or music given a natural language prompt. Diffusion-based frameworks, including the Tango and the AudioLDM series, represent the state-of-the-art in text-to-audio generation. Despite achieving high audio fidelity, they incur significant inference latency due to the slow diffusion sampling process. MAGNET, a mask-based model operating on discrete tokens, addresses slow inference through iterative mask-based parallel decoding. However, its audio quality still lags behind that of diffusion-based models. In this work, we introduce IMPACT, a text-to-audio generation framework that achieves high performance in audio quality and fidelity while ensuring fast inference. IMPACT utilizes iterative mask-based parallel decoding in a continuous latent space powered by diffusion modeling. This approach eliminates the fidelity constraints of discrete tokens while maintaining competitive inference speed. Results on AudioCaps demonstrate that IMPACT achieves state-of-the-art performance on key metrics including Fréchet Distance (FD) and Fréchet Audio Distance (FAD) while significantly reducing latency compared to prior models. The project website is available at \hyperref[https://audio-impact.github.io/]{https://audio-impact.github.io/}.
\end{abstract}

\section{Introduction}
\label{intro}

The text-to-audio generation task aims to synthesize high-quality and high-fidelity audio that aligns semantically with a given textual prompt. This task holds immense potential for applications ranging from audio content creation and video gaming to marketing and advertising.
The current state-of-the-art in text-to-audio generation is represented by the Tango \cite{ghosal2023text, kong24_syndata4genai, majumder2024tango} and AudioLDM \cite{pmlr-v202-liu23f, liu2024audioldm} series, which leverage diffusion-based models to achieve high-quality audio synthesis.
All these models employ computationally heavy network architectures with attention layers as the backbone for their diffusion models. However, this design results in high latency due to slow inference speed, as the iterative denoising steps of the diffusion sampling process combined with the model's complexity significantly increase the time required for generating outputs.

To address the issue of slow inference speed, MAGNET \cite{ziv2024masked}, a masked-based generative model (MGM), utilizes iterative mask-based parallel decoding to achieve efficient audio generation. During inference, the model progressively predicts and refines discrete audio tokens across multiple decoding iterations, leveraging parallelism to predict multiple tokens simultaneously at each step. This parallel decoding strategy not only delivers significantly faster inference compared to traditional autoregressive models like MusicGen \cite{copet2024simple} and AudioGen \cite{kreuk2023audiogen}, but also surpasses the inference speed of diffusion-based models by eliminating the need for time-consuming diffusion sampling.

While MAGNET leverages discrete tokens for efficient and structured audio generation, its performance on text-to-audio generation tasks remains inferior to current state-of-the-art models. 
Given the observed superiority of continuous representations over discrete tokens in tasks such as text-to-image generation \cite{fan2024fluid}, speech large language models \cite{yuan2024continuous}, and automatic speech recognition \cite{xu24d_interspeech}, one intuitive way to enhance the generation performance of MAGNET is to replace its discrete tokens with continuous representations. 
However, based on our preliminary experiments, this intuitive modification resulted in significantly worse performance compared to the original MAGNET model.

Knowing that latent diffusion models (LDMs) are good at modeling continuous representations \cite{ghosal2023text, huang2023make, kong24_syndata4genai, majumder2024tango, pmlr-v202-liu23f, liu2024audioldm, hai2024ezaudio}, we propose to integrate iterative mask-based parallel decoding with LDMs to better model continuous representations for the text-to-audio generation task. 
LDMs require a multi-step diffusion sampling process, which inherently imposes high computational costs and slows inference if used independently. However, by integrating iterative mask-based parallel decoding, we can replace the heavy attention-based layers typically used in LDMs with a lightweight MLP-based diffusion head, substantially reducing sampling time while maintaining audio quality and fidelity.
In addition, we introduce an unconditional pre-training phase before text-conditional training on paired text-audio data, a step shown to be indispensable for this task. Our experimental results confirm that this design enables low-latency inference while preserving high audio fidelity, quality, and text relevance.

In summary, we state our contributions as follows:
\begin{itemize}
    \item We pioneer the use of iterative mask-based parallel decoding on continuous latent representations, powered by LDMs, for text-to-audio generation.
    \item We propose an unconditional pre-training phase preceding conditional training during the MGM training process and demonstrate its effectiveness.
    \item Our model achieves state-of-the-art performance on objective metrics FD and FAD, and subjective evaluations REL and OVL, while remaining competitive with the fastest existing text-to-audio generation model, MAGNET-S, in terms of inference speed.
\end{itemize}

\section{Related Work}
\subsection{Mask-based Generative Models (MGM)}
\label{subsec:mgm}
Mask-based generative modeling (MGM) has emerged as a powerful technique in discrete-token-based sequence modeling to deal with tasks such as audio (Soundstorm \cite{borsos2024soundstorm}, MAGNET \cite{ ziv2024masked}), music (VampNet \cite{garcia2023vampnet}, MAGNET \cite{ziv2024masked}), and image generation (MaskGIT \cite{chang2022maskgit}, MUSE \cite{chang2023muse}, MAGE \cite{li2023mage}).
This approach, which involves masking portions of the input token sequence and training a model to reconstruct the missing information, offers advantages in terms of efficiency and parallelization by employing iterative mask-based parallel decoding. During inference, unlike traditional autoregressive models, such as AudioGen \cite{kreuk2023audiogen}, which generate tokens one at a time, the iterative mask-based parallel decoding process starts with a fully empty sequence and unravels a set of tokens at each decoding iteration to progressively build up a sequence of tokens. The rationale behind this approach is that the generation process starts without prior content. In early iterations, the model has limited context to inform token predictions. As decoding progresses and more tokens are generated, the model gains additional context, enhancing its predictive capabilities in subsequent iterations. 

\subsection{Latent Diffusion Models for Generation}
Latent Diffusion Models (LDMs) are widely employed in text-to-audio tasks due to their ability to operate within a continuous latent space \cite{pmlr-v202-liu23f, liu2024audioldm, ghosal2023text, kong24_syndata4genai, majumder2024tango, hai2024ezaudio}. By encoding audio signals with variational autoencoders (VAEs) \cite{pmlr-v202-liu23f}, LDMs surpass discrete-token-based approaches such as MAGNET \cite{ziv2024masked} in both quality and fidelity. However, due to the iterative nature of diffusion sampling, large LDMs can incur substantial inference overhead. 

In the field of computer vision, MAR \cite{li2024autoregressive}, an MGM-based model, achieved state-of-the-art performance in class-conditional image generation. Unlike earlier MGM-based models such as MaskGIT \cite{chang2022maskgit}, MAGE \cite{li2023mage} and MUSE \cite{chang2023muse} that predict discrete tokens, MAR introduces iterative mask-based parallel decoding directly on continuous representations using LDMs. This choice not only improves image quality and fidelity but also speeds up the diffusion sampling process by replacing heavy attention layers with a lightweight MLP diffusion head.

Inspired by MAR, we present a text-to-audio generation approach that likewise leverages iterative mask-based parallel decoding over continuous latent representations driven by LDMs. We further introduce an unconditional pre-training phase before text-conditional training on paired text-audio data, which is shown to be critical for this task. The results demonstrate that our approach achieves low-latency inference while preserving high audio quality, fidelity, and text relevancy.




\begin{figure*}[t]
    \centering
    \includegraphics[width=17cm]{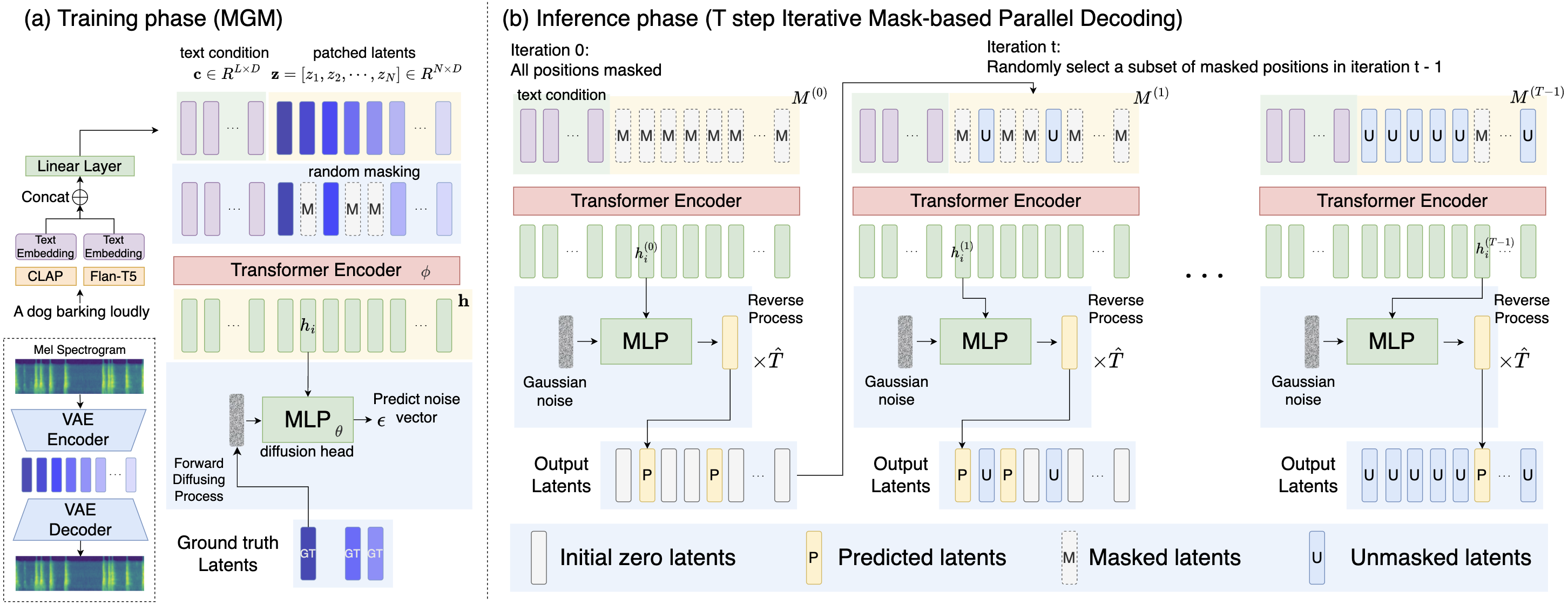}
    \caption{Diagram of our IMPACT framework for the training phase and inference phase.}
    \label{fig:IMPACT_diagram}
\end{figure*}

\section{IMPACT}
IMPACT employs the MGM approach with a conditional LDM. During training, the method first undergoes unconditional pre-training, where no text is used. This phase teaches the model to reconstruct audio latents from partially masked inputs, leveraging large unlabeled datasets to learn the basics of audio generation, which is essential for this task. Next, the model is trained with text conditions by concatenating audio latents with a text condition vector sequence and further encoded by a Transformer-based latent encoder. Here, a small diffusion head predicts the noise used to corrupt masked audio latents, thereby learning to generate audio consistent with text prompts. During inference, the method applies iterative mask-based parallel decoding, starting with a fully empty sequence and gradually generating latents at each iteration with the small diffusion head, saving much time since the diffusion sampling loop operates on a lightweight model.
\subsection{Training Phase}
\subsubsection{Text-conditional Training with MGM}
\label{subsubsec:cmgm}
Figure \ref{fig:IMPACT_diagram}(a) illustrates the MGM training procedure combined with LDMs. Given an audio input, we extract its representation using an audio VAE and arrange it into a sequence of audio latents $\mathbf{z} = [z_1, z_2, \cdots, z_N]$. To train with MGM, a number of $q \cdot N$ latents are masked by a binary mask $M \in \{0, 1\}^N$, where $q$ is the masking percentage factor and $M[i] = 1$ indicates that the $i$-th audio latent is masked. The text condition vector sequence $\mathbf{c}$ is concatenated with the remaining unmasked latents and forwarded to a Transformer-based latent encoder $\text{Enc}_{\phi}$ to produce the hidden representation $\mathbf{h}$, as described by Eq. (\ref{eq:h}): 
\begin{equation}
    \label{eq:h}
    \mathbf{h} = \text{Enc}_{\phi}(\text{concat}(\mathbf{c}, \mathbf{z} \odot \bar{M})), \mathbf{h} \in \mathds{R}^{(L + N) \times D}
\end{equation}
where $\bar{M}$ is the complement of $M$, $L$ denotes the sequence length of $\mathbf{c}$, and $D$ is the encoder embedding dimension.
The goal of our framework is to have a diffusion head to predict these masked audio latents based on input $\mathbf{h}$.
To train such a framework, the latents that were masked are corrupted with noise forming $z_i^{\hat{t}} = \sqrt{\bar{\alpha}_{\hat{t}}} z_i + \sqrt{1 - \bar{\alpha}_{\hat{t}}} \epsilon$ based on the closed-form solution of the forward diffusing process \cite{ho2020denoising, nichol2021improved}, where $\epsilon \in \mathds{R}^D$ denotes a noise vector sampled from the multivariate normal distribution $\mathcal{N}(0, \mathbf{I})
$, $\bar{\alpha}_{\hat{t}}$ is the noise schedule, and $\hat{t}$ represents the time step of the noise schedule, also known as the diffusion step. A diffusion head $\epsilon_{\theta}$, parameterized by $\theta$, takes $h_i$ as a condition to predict the noise $\epsilon$ used to corrupt the latents. Eq. (\ref{eq:diff_loss}) shows the training objective, 
\begin{equation}
    \label{eq:diff_loss}
     \arg\min_{\{\phi,\theta\}} \sum_{\{i \mid M[i] = 1\}} \left\lVert \epsilon - \epsilon_{\theta}(z_i^{\hat{t}} \mid \hat{t}, h_i) \right\rVert^2
\end{equation}
where $\{i\mid M[i] = 1 \}$ is the set of indices $i$ of $M$ such that $M[i] = 1$, representing the positions of $\mathbf{z}$ that are masked.
$\hat{t}$ denotes the integer-valued diffusion step, sampled from the interval $[0, \hat{T}_{\text{max}} ]$.
$\{\phi, \theta\}$ denotes the set of parameters of the Transformer-based latent encoder and the diffusion head. Both modules are jointly optimized with the objective shown in Eq. (\ref{eq:diff_loss}).


\subsubsection{Unconditional Pre-training}
\label{subsubsec:uncond}
Unconditional pre-training is performed in a similar manner mentioned in Section \ref{subsubsec:cmgm} but without text conditions. It performs mask-based generative modeling without the text condition vector sequence $\textbf{c}$. This process serves as a preliminary training phase before we have the model to learn how to follow text descriptions during audio generation. The benefits of this come in twofold: (1) it allows the model to first gain generative abilities for audio generation and relieves the burden of having to learn both generation and text relevancy in the same phase, (2) it allows us to utilize unpaired data since some audio datasets like AudioSet do not have text descriptions for all audio samples.
\subsection{Inference Phase}
Figure \ref{fig:IMPACT_diagram}(b) illustrates the iterative mask-based parallel decoding method performed during the inference phase. The decoding process starts from a fully empty sequence $\mathbf{z}^{(0)}$ and a full mask $M^{(0)} = [1, 1, \cdots, 1] \in \{0, 1\}^N$. 
The decoding process is composed of three main stages: random position selection, mask scheduling, and diffusion sampling.
We elaborate on them in the following sections.
\subsubsection{Random Position Selection}
\label{subsubsec:mask_strategy}
Previous mask-based generative models for audio, such as Soundstorm, VampNet, and MAGNET, rely on discrete token representations and selectively predict token positions with low confidence scores at each decoding iteration. In contrast, our approach operates on continuous representations, making it infeasible to compute confidence scores. Therefore, at each decoding iteration $t$,  we randomly select a subset of unpredicted positions in $\mathbf{z}^{(t)}$. We denote the set of indices of the positions to be predicted at each decoding iteration $t$ as $M^{(t)}_{\text{pred}}$. These selected positions are generated in parallel rather than sequentially to reduce inference time.

\subsubsection{Mask Scheduling}
\label{subsubsec:mask_sched}
At each decoding iteration $t$, a masking scheduler determines the number of latents to remain masked, denoted as $\mu^{(t)}$, based on a fraction $\gamma^{(t)}$ of the $N$ latents: $\mu^{(t)} = \left\lfloor \gamma^{(t)} \cdot N \right\rfloor.$
The fraction $\gamma^{(t)}$ decreases over iterations following a predefined masking schedule, commonly defined as $\gamma^{(t)} = \cos\left( \frac{\pi}{2} \cdot \frac{t}{T} \right)$, where $T$ is the total number of decoding iterations. This cosine masking schedule ensures that the number of masked latents decreases with each iteration, progressively increasing the amount of information available to the model for the next decoding iteration. The details of masking implementation are listed in Appendix \ref{app:mask_stra} and \ref{app:mask_sched}.

\subsubsection{Diffusion Sampling}
At each decoding iteration $t$, the model generates latents for the positions specified by $M^{(t)}_{\text{pred}}$. Given the hidden representation $\mathbf{h}^{(t)}$ produced by the Transformer-based latent encoder, latents $z_i = z_i^0$ are sampled by following the reverse process \cite{ho2020denoising, nichol2021improved} shown in Eq. (\ref{eq:reverse}), 
\begin{equation}
    \label{eq:reverse}
    z^{\hat{t}-1}_i=\frac{1}{\sqrt{\alpha_{\hat{t}}}}\left(z^{\hat{t}}_i-\frac{1-\alpha_{\hat{t}}}{\sqrt{1-\overline{\alpha}_{\hat{t}}}}\epsilon_{\theta}(z^{\hat{t}}_i\ |\ \hat{t}, h^{(t)}_i)\right)+\sigma_{\hat{t}}\delta,
\end{equation}
where $\sigma_{\hat{t}}$ denotes the noise level at diffusion sampling step $\hat{t}$ and $\delta$ denotes a vector drawn from $\mathcal{N}(0, \mathbf{I})$. Note that the diffusion sampling step $\hat{t}$  is distinct from the decoding iteration $t$.

During the iterative decoding process, classifier free guidance (cfg) \cite{ho2021classifierfree} is adopted to balance the text relevancy and audio fidelity. More details of cfg are elaborated in Appendix \ref{app:cfg}.
\section{Experimental Setup}
\subsection{Dataset Configurations}
Multiple audio datasets are involved in this work. 
Specifically, we employ the AudioCaps (AC) \cite{kim-etal-2019-audiocaps} training split, which contains 145 hours of audio, and a combined dataset of AudioCaps (AC) and WavCaps (WC) \cite{10572302}, totaling 1200 hours of audio.
Although AudioSet (AS) \cite{45857} is currently the largest audio dataset which has about 5500 hours of audio data, since most of the audio samples in AS do not have text descriptions, this dataset is only used for unconditional pre-training. 
For evaluation, we evaluate our text-to-audio generation model on the AC evaluation set. There are 5 text descriptions for each sample in AC, and we follow AudioLDM by randomly selecting one text description as the text condition.
For data preprocessing, we follow AudioLDM's recipe by segmenting each audio sample into 10 seconds and extracting the Mel spectrogram. 
\subsection{Implementation Details and Model Configurations}
Our IMPACT model is composed of three main components, the VAE module, the Transformer-based latent encoder, and the diffusion head.
For the VAE module, we directly adopt the VAE of AudioLDM to extract raw audio latents $\mathbf{z}^\prime \in \mathds{R}^{H \times W \times \text{ch}}$ from Mel spectrograms, where $H = 256$, $W = 16$ and $\text{ch} = 8$. A patching operation with factor $p = 4$ is performed to reduce the height and width dimension of $\mathbf{z}^\prime$ into $\mathbf{z}^{\prime\prime} \in \mathds{R}^{\frac{H}{p} \times \frac{W}{p} \times \text{ch}*p^2}$. The patched latents $\mathbf{z}^{\prime\prime}$ are then flattened and projected to an embedding dimension of $D$ resulting in $\mathbf{z} \in \mathds{R}^{N \times D}$, where $N = \frac{H}{p} \times \frac{W}{p} = 256$. The patching operation reduces the sequence length to $N$ and hence reduces computation during the generation process. 
For the text condition, concatenating CLAP \cite{wu2023large} and Flan-T5 \cite{chung2024scaling} embeddings on the time dimension results in a length of $L = 78$. The embedding dimension of the text conditioning vector sequence is also projected to $D$ resulting in $\mathbf{c} \in \mathds{R}^{L \times D}$.
In this work, we develop two configurations of the IMPACT model, differing in the size of the Transformer-based latent encoder $\text{Enc}$: a base configuration and a large configuration. The base configuration uses an embedding dimension $D$ of 768 and incorporates 24 transformer layers in the latent encoder. In contrast, the large configuration increases the embedding dimension to $D = 1024$ and employs 32 transformer layers in the encoder.
For the diffusion head, we employ a multi-layer perceptron (MLP) adopted from MAR \cite{li2024autoregressive} to effectively incorporate conditioning information from both diffusion steps and conditioning vectors $h_i$. No heavy parameterized attention-based layers are used in the diffusion head to avoid high computational costs during the diffusion loop of the diffusion sampling process at the inference phase. 
In both the base and large configurations, the diffusion head remains identical, as scaling it up would substantially degrade inference speed.
During training, the maximum number of diffusion steps $\hat{T}_{\text{max}}$ is set to 1000. During inference, the total number of diffusion sampling steps $\hat{T}$ is set to 100 unless specified explicitly. 
Additional information on the Transformer-based latent encoder and diffusion head can be found in Appendix \ref{app:trans_enc} and \ref{app:diff_head}.

During text-conditional training, we set the masking percentage factor $q = 0.7$. 
For all model training, we adopt the AdamW optimizer and set the learning rate to $5\mathrm{e}{-5}$. 
For inference, by default, the total number of decoding iterations $T$ is set to 64 unless otherwise specified. For classifier free guidance, we list the details in Appendix \ref{app:cfg}.

\begin{table*}[h]
\footnotesize
\setlength\tabcolsep{3.0 pt}
\caption{System level comparison of various text-to-audio generation models and their performance on objective and subjective metrics. Models publicly available were inferenced by us with parameters adjusted to achieve the best possible performance. Results for models that are not publicly available (marked with $^\star$) are directly reported from their original papers or from other existing works that have documented them. The notations ``pt. data" and ``tc. data" represent the training data durations, measured in hours, for pre-training and text-conditional training, respectively. For IMPACT models ``pt. data" specifically refers to the duration of training data for unconditional pre-training. Detailed information on the training data is listed in Appendix \ref{app:data}. The abbreviation ``diff.'' denotes the number of diffusion sampling steps used for inference. The abbreviation ``Lat." represents the latency of generating a batch of 8 audios measured in seconds. 
Best performance values are marked in bold. Second-best performance values are underscored.}
\label{tab:main_result}
\vskip 0.1in
\centering
\begin{tabular}{lccc|ccccc|cc|cc}
\toprule
\textbf{AudioCaps} & \textbf{pt. data} & \textbf{tc. data} & \textbf{\# para} & \textbf{FD $\downarrow$} & \textbf{FAD $\downarrow$} & \textbf{KL $\downarrow$} & \textbf{IS $\uparrow$} & \textbf{CLAP $\uparrow$} & \textbf{REL $\uparrow$} & \textbf{OVL $\uparrow$} & \textbf{diff.} & \textbf{Lat. 
$\downarrow$}\\
\midrule
Ground Truth & - & - & - & - & - & - & - & 0.373 & 4.43 & 3.57 & - & -\\
\midrule
AudioGen$^\star$ & - & $\approx$ 4000 & 1B & - & 1.82 & 1.69 & - & - & - & - & - & -\\
AudioGen v2 & - & $\approx$ 4000 & 1.5B & 16.51 & 2.11 & 1.54 & 9.64 & 0.315 & - & - & - & 37.2 \\
Tango & - & 145 & 866M & 24.42 & 1.73 & 1.27 & 7.70 & 0.313 & - & - & 200 & 182.6 \\
Tango-full-ft & $\approx$ 3333 & 145 & 866M & 18.93 & 2.19 & 1.12 & 8.80 & 0.340 & - & - & 200 & 181.6 \\
Tango-AF\&AC-FT-AC & $\approx$ 400 & 145 & 866M & 21.84 & 2.35 & 1.32 & 9.59 & 0.343 & - & - & 200 & 182.6 \\
Tango 2 & $\approx$ 3333 & $\approx$ 80 & 866M & 20.66 & 2.63 & 1.12 & 9.09 & 0.375 & 4.13 & 3.37 & 200 & 182.3\\
EzAudio-L (24kHz) & $>$ 5500 & 145 & 596M & 15.59 & 2.25 & 1.38 & \underline{11.35} & \textbf{0.391} & 4.05 & 3.44 & 50 & 29.1 \\
EzAudio-XL (24kHz) &  $>$ 5500 & 145 & 874M & 14.98 & 3.01 & 1.29 & \textbf{11.38} & \underline{0.387} & 4.00 & 3.35 & 50 & 40.2\\
MAGNET-S & - & $\approx$ 4000 & 300M & 23.02 & 3.22 & 1.42 & 9.72 & 0.287 & 3.83 & 2.84 & - & \textbf{6.9} \\
MAGNET-L & - & $\approx$ 4000 & 1.5B & 26.19 & 2.36 & 1.64 & 9.10 & 0.253 & - & - & - & 24.8 \\
Make-an-Audio 2 & - & 3700 & 160M & 16.23 & 2.03 & 1.29 & 9.95 & 0.345 & - & - & 100 & 15.9 \\
AudioLDM2-AC-large$^\star$ & - & 145 & 712M & - & 1.42 & \textbf{0.98} & - & - & - & - & - & -\\
AudioLDM2-full & - & 29510 & 346M & 32.14 & 2.17 & 1.62 & 6.92 & 0.273 & 3.74 & 3.19 & 200 & 96.1 \\
AudioLDM2-full-large & - & 1150k & 712M & 33.18 & 2.12 & 1.54 & 8.29 & 0.281 & - & -  & 200 & 195.7\\
\midrule
\midrule
IMPACT base, dec iter 32 & 1200 & 1200 & 193M &  \underline{14.90} & \textbf{1.07} & \underline{1.05} & 10.06 & 0.364 & - & - & 100 & \underline{11.2} \\ 
IMPACT base, dec iter 64 & 5500  & 1200 & 193M & \textbf{14.72} & \underline{1.13} & 1.09 & 10.03 & 0.353 & \underline{4.31} & \textbf{3.51} & 100 & 22.2 \\
IMPACT large, dec iter 64 & 5500  & 1200 & 427M & \textbf{14.72} & 1.17 & 1.07 & 10.53 & 0.364 & \textbf{4.39} & \underline{3.47} & 100 & 28.4\\

\bottomrule
\end{tabular}
\vskip -0.1in
\end{table*}

\subsection{Evaluation Metrics}
\label{subsec:eval_metric}

In this work, generated audios are evaluated on five objective metrics: Fréchet Distance (FD) \cite{heusel2017gans}, Fréchet Audio Distance (FAD) \cite{kilgour19_interspeech}, Kullback–Leibler divergence (KL), Inception Score (IS) \cite{salimans2016improved}, and Contrastive Language-Audio Pretraining (CLAP). The meanings and implementation details of these metrics are elaborated in Appendix \ref{app:objective_metrics}.

For subjective evaluation, following a similar approach to Tango \cite{ghosal2023text}, we assessed 30 generated audio samples based on text relevance (REL) and overall quality (OVL) but used a 1-to-5 rating scale instead of a 1-to-100 scale. Each sample was rated by at least 10 participants. 
See Appendix \ref{app:human_eval} for more details on the rating platform.

To evaluate inference speed, we measure the latency, also referred to as inference time, reported in seconds for generating a batch of audio samples on a single Tesla V100 GPU with 32GB VRAM.

\section{Results}
\label{exp}
We evaluate our proposed text-to-audio framework across three key aspects. First, Table \ref{tab:main_result} reports results on the AudioCaps benchmark, using the objective and subjective metrics defined in Section \ref{subsec:eval_metric}, and contrasts our models against the current state of the art, while Table \ref{tab:impact_results} presents ablations over key training configurations. Second, Figure \ref{fig:all_vs_niter} and Table \ref{tab:diff_step} analyze the trade-off between objective performance and latency, focusing on the two primary parameters affecting inference time: the number of decoding iterations and diffusion steps. Finally, Figure \ref{fig:latency_througput} compares its latency and throughput with MAGNET-S, the fastest existing text-to-audio model.

\subsection{Overall System-level Performance}
Table \ref{tab:main_result} shows the system-level performance of text-to-audio generation on the AudioCaps evaluation set measured in both objective and subjective metrics. We compare our proposed models with current state-of-the-art models AudioGen \cite{kreuk2023audiogen}, Tango \cite{ghosal2023text}, Tango-full-ft \cite{ghosal2023text}, Tango-AF\&AC-FT-AC \cite{kong24_syndata4genai}, Tango 2 \cite{majumder2024tango}, EzAudio \cite{hai2024ezaudio}, MAGNET \cite{ziv2024masked}, Make-an-Audio 2 \cite{huang2023make}, and AudioLDM2 \cite{liu2024audioldm}.
As shown in Table \ref{tab:main_result}, our proposed IMPACT model outperforms current state-of-the-art models on the FD and FAD metrics. Specifically, while EzAudio-XL achieves the lowest FD score of 14.98 among existing baselines, Tango attains an FAD score of 1.73, our proposed IMPACT base and large models surpass these results on the respective metrics.
For the KL metric, Tango-full-ft and Tango 2 both achieve a KL score of 1.12. In contrast, our proposed IMPACT base and large models attain better KL scores, falling only slightly behind the non-public model AudioLDM2-AC-large.
Regarding the CLAP metric, our IMPACT models achieve a score of 0.364, slightly lower than Tango 2's 0.375, which may be attributed to Tango 2's use of the Direct Preference Optimization (DPO) \cite{rafailov2023direct} method for training. 
Note that EzAudio poses a huge gap in CLAP score compared to all other text-to-audio models. This is likely due to applying a data filtering method derived from CapFilt \cite{li2022blip} to discard audio-text pairs with CLAP scores lower than a certain threshold.
Nonetheless, subjective evaluations indicate that our IMPACT models surpass Tango 2 and EzAudio-XL in text relevancy (REL), demonstrating their competitiveness in ensuring the generated audio closely corresponds to the provided text descriptions from human perspectives. 

The last column of Table \ref{tab:main_result} compares the inference speed, also known as latency, of IMPACT with that of all other baseline models, each configured using hyperparameters that yield their overall optimal performance on objective metrics. While our IMPACT model with 32 decoding iterations achieves the second-lowest latency, just behind MAGNET-S, it significantly outperforms MAGNET-S across multiple objective metrics. Ablation studies on the two key factors that influence latency, the number of decoding iterations and diffusion steps, are presented in Section \ref{subsec:infer_config}.

\subsection{Training Configurations}
\subsubsection{Unconditional Pre-training}
We investigate the benefits of unconditional pre-training by comparing three IMPACT models in Table \ref{tab:impact_results}: (a) trained exclusively with text-conditional training, and (b) and (c) additionally pre-trained unconditionally using different data scales. 
Compared to model (a), model (c) exhibits consistent gains across objective metrics.
This improvement can be attributed to the unconditional pre-training phase, which enables the model to ensure the audio quality and fidelity for generation. The subsequent text-conditional training then aligns the model to follow text descriptions more effectively while further enhancing audio quality and fidelity.
Furthermore, although increasing the pre-training data from model (b) to model (c) slightly degrades KL, IS, and CLAP, subjective evaluations consistently favor model (c), underscoring the value of larger-scale data for unconditional pre-training.
These findings demonstrate the importance of the unconditional pre-training phase and how it contributes to the performance of IMPACT models.

\subsubsection{Text-conditional Training Data}
By comparing models (c) and (d) in the IMPACT base configuration, we observe notable improvements in key metrics as the amount of text-conditioning training data is increased. Both models were unconditionally pre-trained with 5500 hours of data; however, model (d) was conditionally trained on 145 hours of data, while model (c) was conditionally trained on 1200 hours of data. This increase in text-conditioning training data led to substantial performance gains across several metrics. Model (c) achieved a lower FAD of 1.13 compared to 1.38 for model (d). Additionally, model (c) demonstrated a reduced KL divergence, scoring 1.09 versus 1.16 for model (d). Model (c) also surpassed model (d) in CLAP score, achieving 0.353 compared to 0.340. These findings confirm that increasing text-conditional training data enhances the performance of generated audio on objective metrics.

To replicate the training dataset configuration of Tango-full-ft \cite{ghosal2023text}, which was initially pre-trained on a large-scale dataset and subsequently fine-tuned on AC, we derive model (c') by performing text-conditional training on model (c) again using the AC training set. This additional training improves the performance on FD and FAD metrics, reflecting improved audio fidelity, but modestly degrades IS and CLAP scores, suggesting a potential trade-off between audio fidelity and quality, and also semantic consistency between the generated audio and the provided text prompt. 

\begin{table*}[t]
\footnotesize
\setlength\tabcolsep{3.0 pt}
\caption{Model performance comparisons between various IMPACT models on objective and subjective metrics. The $^\dagger$ notation specifies that model (c') is trained with text-conditional training twice. Detailed latency values of baseline models and IMPACT are listed in Appendix \ref{app:latency}.}
\label{tab:impact_results}
\vskip 0.1in
\centering
\begin{tabular}{llccc|ccccc|cc|c}
\toprule
&\textbf{AudioCaps} & \textbf{pt. data} & \textbf{tc. data} & \textbf{\# para} & \textbf{FD $\downarrow$} & \textbf{FAD $\downarrow$} & \textbf{KL $\downarrow$} & \textbf{IS $\uparrow$} & \textbf{CLAP $\uparrow$} & \textbf{REL $\uparrow$} & \textbf{OVL $\uparrow$} & \textbf{Lat. 
$\downarrow$}\\
\midrule
(a)& IMPACT base, dec iter 64 & - & 1200 & 193M & 14.86 & 1.35 & 1.17 & 9.75 & 0.346 & - & - & 22.2 \\ 
(b)& IMPACT base, dec iter 64 & 1200 & 1200 & 193M & 15.36 & 1.13 & \underline{1.04} & 10.37 & 0.361 & 4.15 & 3.45 & 22.2\\ 
(b')& IMPACT base, dec iter 32 & 1200 & 1200 & 193M &  14.90 & \textbf{1.07} & 1.05 & 10.06 & 0.364 & - & - & 11.2 \\ 
(c)& IMPACT base, dec iter 64 & 5500  & 1200 & 193M & 14.72 & 1.13 & 1.09 & 10.03 & 0.353 & \underline{4.31} & \textbf{3.51} & 22.2 \\
(c')& IMPACT base, dec iter 64 & 5500  & 1200$^\dagger$ & 193M & \textbf{13.86} & \underline{1.12} & 1.11 & 9.41 & 0.347& - & - & 22.2\\
(d)& IMPACT base, dec iter 64  & 5500 & 145 & 193M & 15.18 & 1.38 & 1.16 & 9.19 & 0.340 & - & - & 22.2\\
\midrule 
(e)& IMPACT large, dec iter 64 & 5500 & 145 & 427M & \underline{14.50} & 1.50 & 1.13 & 9.53 & 0.343 & - & - & 28.4\\
(f)& IMPACT large, dec iter 64 & 5500  & 1200 & 427M & 14.72 & 1.17 & 1.07 & 10.53 & 0.364 & \textbf{4.39} & \underline{3.47} & 28.4\\

\bottomrule
\end{tabular}
\vskip -0.1in
\end{table*}

\begin{table}[ht]
\setlength\tabcolsep{3.0 pt}
\caption{Performance comparisons of IMPACT models training with or without (w/o) CLAP embeddings.}
\label{tab:clap_ablation}
\vskip 0.15in
\renewcommand{\arraystretch}{0.4}
\centering
\begin{tabular}{lcccc}
\toprule
\textbf{dec iter 64} & \textbf{FAD ↓} & \textbf{KL ↓} & \textbf{IS ↑} & \textbf{CLAP ↑} \\
\midrule
IMPACT model (c) & 1.13 & 1.09 & 10.03 & 0.353 \\
\quad w/o CLAP & 1.13 & 1.10 & 10.22 & 0.352 \\
\midrule
IMPACT model (d) & 1.38 & 1.16 & 9.19 & 0.340 \\
\quad w/o CLAP & 1.49 & 1.10 & 9.19 & 0.344 \\

\bottomrule
\end{tabular}
\vskip -0.1in
\end{table}

\subsubsection{CLAP module}
In this work, we adopt two text encoders, CLAP and FlanT5, to encode text into conditions. To assess the contribution of CLAP-based conditioning, in Table \ref{tab:clap_ablation}, we conduct ablation studies by removing the CLAP encoder during training while continuing to evaluate text relevance using CLAP scores. The results show that even without the CLAP module, the model’s performance remains relatively stable across key metrics, with only minor differences observed. For example, the FAD and KL scores remained stable for IMPACT model (c), and the CLAP score showed only a negligible decrease from 0.353 to 0.352. These findings suggest that the model’s performance, particularly in terms of semantic alignment measured by the CLAP score, is robust to the removal of the CLAP module, indicating that the model can effectively learn text-audio alignment without explicit CLAP-based input, relying solely on FlanT5 embeddings.
Similarly, for IMPACT model (d), removing CLAP slightly increased the FAD from 1.38 to 1.49 and decreased the KL score from 1.16 to 1.10. However, the CLAP score still remained similar while removing the CLAP embeddings, indicating that semantic alignment between generated audio and text was not significantly impacted. 

\begin{table}[ht]
    \caption{Performance of IMPACT with single-pass decoding and 32-iteration decoding. Masking percentage factor $q$ is set to $1.0$ for training to yield the best performance for the single-pass model.}
    \label{tab:single_vs_32}
    \vskip 0.1in
    \centering
    \begin{tabular}{lcccc}
        \toprule
        & \textbf{FAD} $\downarrow$ & \textbf{KL} $\downarrow$ & \textbf{IS} $\uparrow$ & \textbf{CLAP} $\uparrow$ \\
        \midrule
        single-pass & 12.26 & 2.57 & 2.84 & 0.125 \\ 
        dec iter 32 & \textbf{1.07} & \textbf{1.05} & \textbf{10.06} & \textbf{0.364} \\
        \bottomrule
    \end{tabular}
    \vskip -0.1in
\end{table}

\begin{table}[ht]
    \caption{Performance of IMPACT model (b) using different diffusion steps $\hat{T}$ for inference. Total decoding iterations set to 64.}
    \label{tab:diff_step}
    \vskip 0.05in
    \centering
    \begin{tabular}{ccccc|c}
        \toprule
        $\hat{T}$ & \textbf{FAD $\downarrow$} & \textbf{KL $\downarrow$} & \textbf{IS $\uparrow$} & \textbf{CLAP $\uparrow$} & \textbf{Lat. 
$\downarrow$} \\
        \midrule
        50  & 1.57 & 1.15 & 10.02 & 0.342 & \textbf{12.1}\\ 
        100 & \textbf{1.13} & \textbf{1.04} & \textbf{10.37} & 0.361 & 22.2 \\ 
        150 & 1.30 & 1.06 & 10.31 & 0.363 & 31.6\\ 
        200 & 1.19 & 1.06 & 10.36 & \textbf{0.364} & 41.9\\ 
        \bottomrule
    \end{tabular}
    \vskip -0.1in
\end{table}
\begin{figure*}[t]
    \vskip 0.1in
    \centering
    \includegraphics[width=17cm]{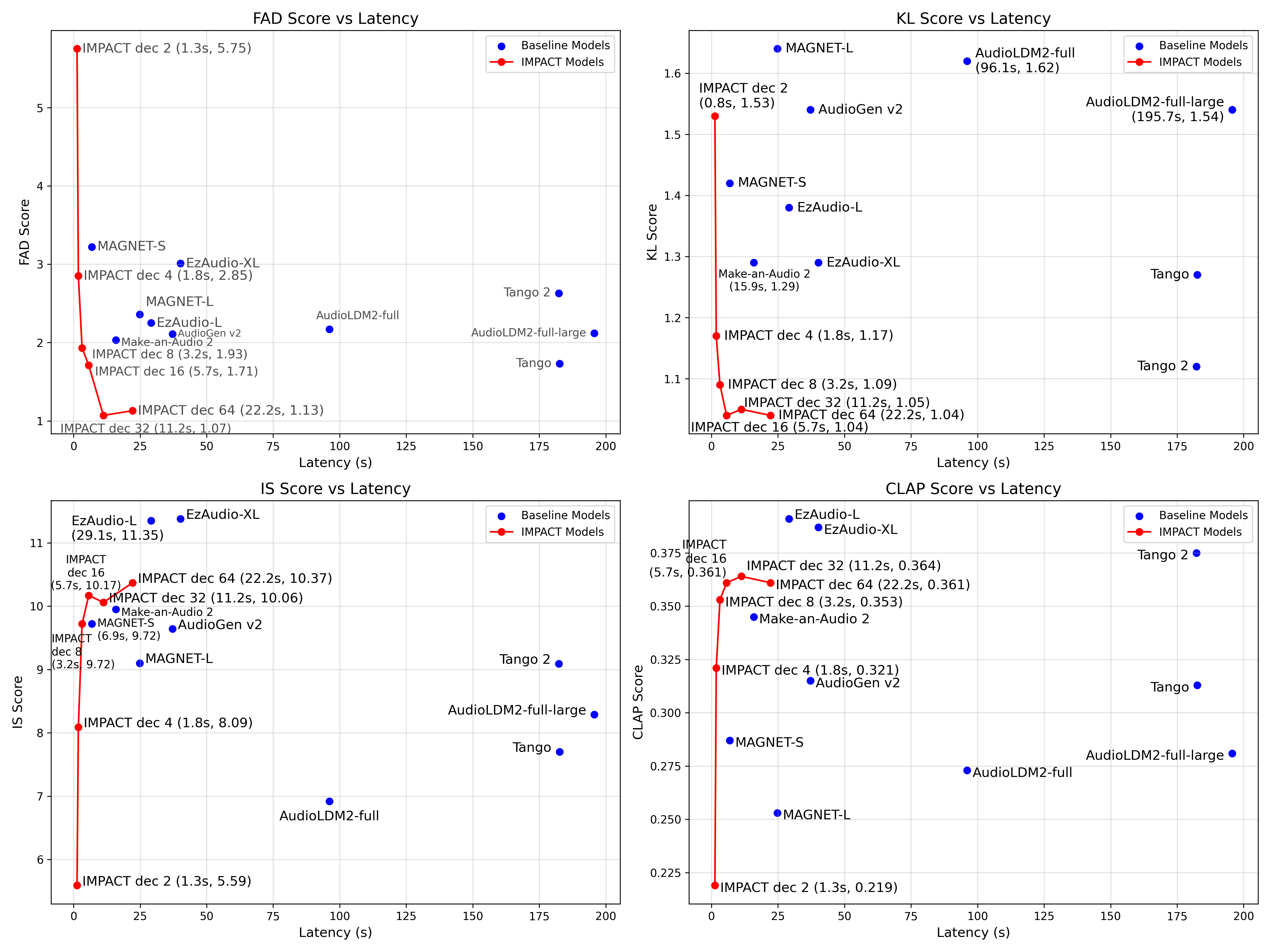}
    \vskip -0.1in
    \caption{Performance of the baseline models and our proposed IMPACT model (b) with varying decoding iterations (dec iter), visualized by plotting objective metrics (FAD, IS, KL, and CLAP) against latency. Data points of the IMPACT model are plotted in red curves. 
    }
    \label{fig:all_vs_niter}
    \vskip -0.1in
\end{figure*}
\begin{figure}[ht]
    \vskip -0.05in
    \centering
    \includegraphics[width=8cm]{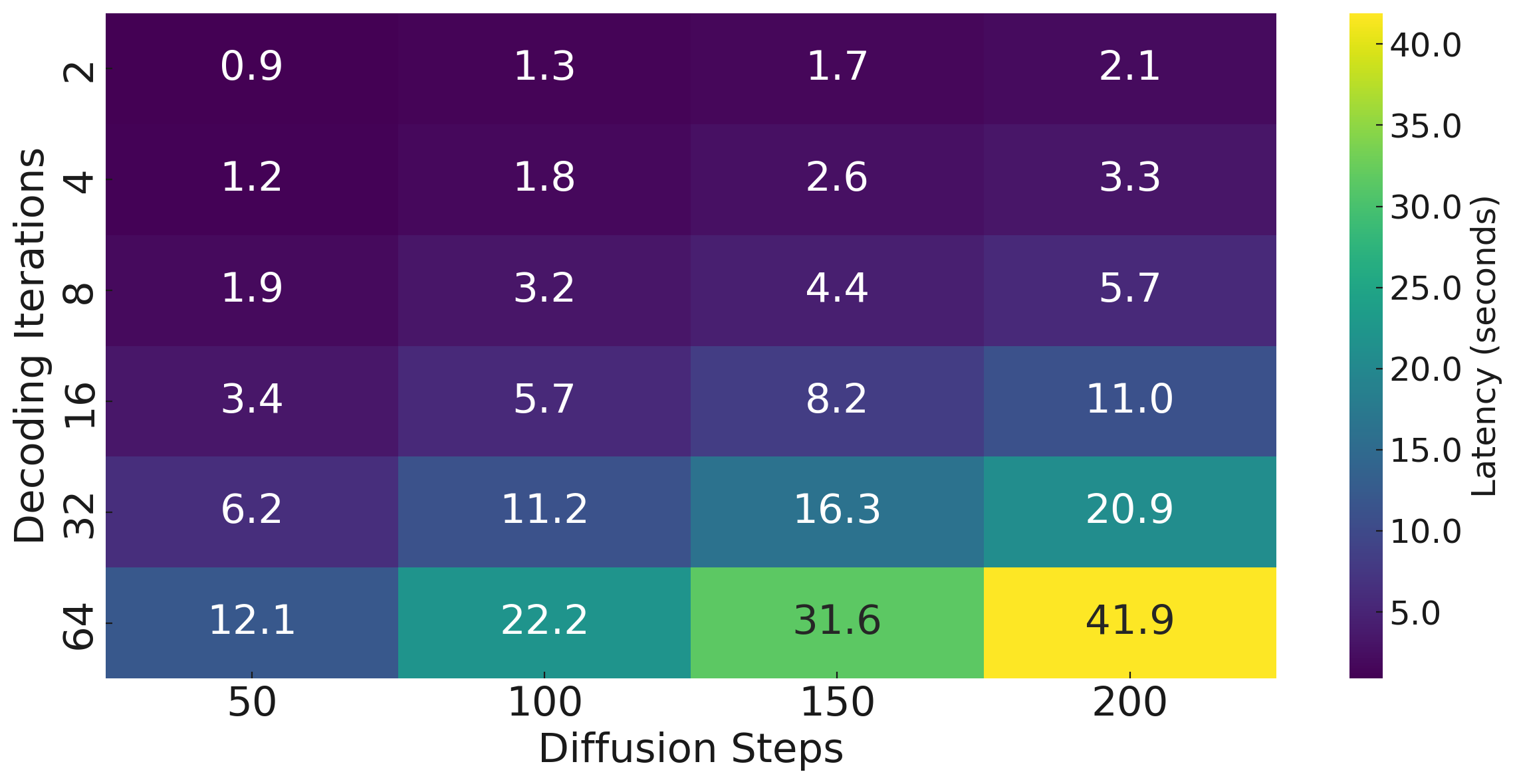}
    \vskip -0.1in
    \caption{Heatmap visualizing latency measurements for IMPACT model (b) under varying decoding iterations and diffusion steps at a batch size of 8. Latency is depicted by color intensity, with brighter colors indicating higher values. Detailed objective performance values are provided in Appendix \ref{app:dec_iter_diff_step}.}
    \label{fig:dec_iter_diff_step_latency}
    \vskip -0.1in
\end{figure}

\subsection{Inference Configurations}
\label{subsec:infer_config}
\subsubsection{Decoding Iterations}
\label{subsubsec:dec_iter}
Table \ref{tab:single_vs_32} studies the effectiveness of iterative decoding. The single-pass model, trained to generate audio latents in a single-pass, performs much worse than the model with 32 steps of iterative decoding. This showcases the importance of gradually generating the audio latents in an iterative manner since further iterations can leverage the previously generated latents as the condition for generation.
Figure \ref{fig:all_vs_niter} shows the objective performance versus latency (inference time) of the baseline models and IMPACT model (b) using different decoding iterations during the inference phase.
With 16 decoding iterations, IMPACT surpasses all publicly available baselines on the FAD metric, while only 8 iterations are sufficient to exceed these baselines in terms of the KL score. For IS scores, we observe an upward trend as the number of decoding iterations increases. For KL and CLAP scores, the gain of performance with 16 decoding iterations or more is little. 
When using up to 16 decoding iterations, we observe a strong correlation between performance and decoding iterations across all four objective metrics, indicating a trade-off between audio quality, fidelity, and inference speed, as more decoding iterations require more inference time. 
Notably, with 16 iterations, IMPACT not only surpasses all baselines on FAD and KL but also outperforms most of them on IS, all within a latency of just 5.7 seconds, which is significantly lower than any of the baseline models.

\subsubsection{Diffusion Steps}
\label{subsubsec:diff_step}
Table \ref{tab:diff_step} analyzes the effect of using different numbers of diffusion steps for inference. For diffusion steps 100, 150, and 200, the KL, IS, and CLAP scores are similar. For FAD, KL and IS scores, the best performance happens when using 100 steps. Given the fact that more diffusion steps result in higher latency, we conclude that using 100 diffusion steps is the most suitable parameter to achieve high performance on the objective metrics and to have low latency.

\subsubsection{Optimal Parameters}
Figure \ref{fig:dec_iter_diff_step_latency} analyzes how decoding iterations and diffusion steps together affect the latency of IMPACT. It is observed that the impact of increasing diffusion steps on latency becomes more severe at higher decoding iterations. 
The latency increase due to decoding iterations is more tolerable at lower diffusion steps.
Based on the results of Sections \ref{subsubsec:dec_iter} and \ref{subsubsec:diff_step}, we conclude that using 100 diffusion steps along with 16 to 64 decoding iterations is the optimal range of parameters for our IMPACT models.

\begin{figure*}[t]
    \vskip 0.1in
    \centering
    \includegraphics[width=17.2cm]{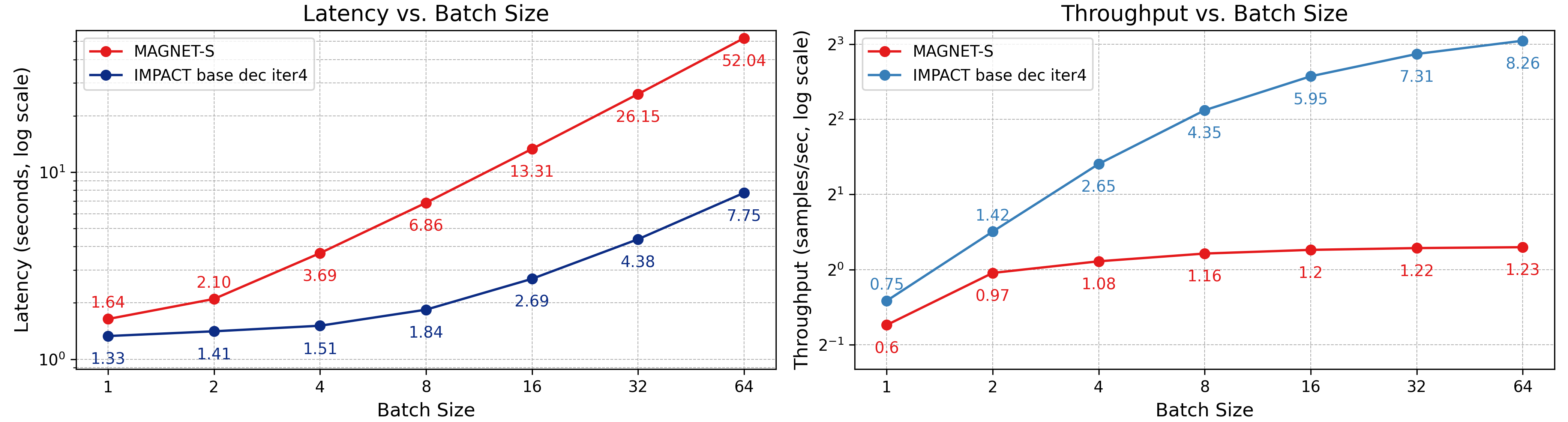}
    \vskip -0.1in
    \caption{Batch latency and throughput comparisons between MAGNET-S and our IMPACT base model (b) with 4 decoding iterations.}
    \label{fig:latency_througput}
\end{figure*}
\subsection{Latency and Throughput Compared to MAGNET}
IMPACT can flexibly adjust the number of decoding iterations to balance objective performance metrics and inference efficiency. To analyze this trade-off against the fastest existing baseline, Figure \ref{fig:latency_througput} presents the latency and throughput of MAGNET-S and our IMPACT base model with 4 decoding iterations across various batch sizes. Here, we define latency as the time in seconds to generate each batch of audio, and throughput as the number of audio samples generated per second.
As reported in Table \ref{tab:niter4_magnets}, even with just 4 decoding iterations, IMPACT already surpasses MAGNET-S in FAD, KL, and CLAP. IMPACT achieves faster inference than MAGNET-S, even with looping through the diffusion sampling process, due to its efficiency in reaching high objective performance with fewer decoding iterations. In contrast, MAGNET requires decoding across four levels of codebooks, which needs 50 iterations in total.
Furthermore, when examining performance on batch sizes ranging from 1 to 64, IMPACT demonstrates consistently lower inference time and higher throughput, indicating that it scales better with larger batch sizes. For example, at a batch size of 64, IMPACT completes inference in 7.75 seconds with a throughput of 8.26 samples per second, compared to MAGNET-S’s 52.04 seconds and 1.23 samples per second, respectively. This improvement highlights IMPACT’s scalability and speed, making it a compelling option for applications that demand both high throughput and low latency on single-GPU systems, while also offering competitive or superior output quality.

\begin{table}[ht]
    \setlength\tabcolsep{4.0 pt}
    \caption{Comparison between MAGNET-S and IMPACT. The number of diffusion steps is set to 100 for IMPACT model (b).}
    \label{tab:niter4_magnets}
    \vskip 0.1in
    \centering
    \begin{tabular}{lcccc}
        \toprule
        & dec iter & \textbf{FAD} $\downarrow$ & \textbf{KL} $\downarrow$ & \textbf{CLAP} $\uparrow$ \\
        \midrule
        MAGNET-S & 50   & 3.22 & 1.42 & 0.287 \\
        IMPACT model (b) & 4  & \textbf{2.85} & \textbf{1.17} & \textbf{0.321} \\
        \bottomrule
    \end{tabular}
\end{table}

\section{Conclusions}
In this paper, we introduce IMPACT for text-to-audio generation, which combines the advantages of iterative mask-based parallel decoding and continuous latent representations through LDMs. By leveraging continuous latents with LDMs, IMPACT overcomes the limitations of discrete-token-based methods, offering superior audio fidelity and semantic alignment. Moreover, its mask-based decoding mechanism and adoption of a small diffusion head for generation ensure efficient inference, achieving faster inference speed than multiple baseline models and competitive latency compared to the fastest text-to-audio model, MAGNET-S. Apart from these methods, we propose unconditional pre-training and demonstrate the importance of it to achieve state-of-the-art performance on objective metrics such as FD and FAD. Our extensive experiments on the AudioCaps evaluation set highlight IMPACT's ability to balance audio quality, fidelity, text relevancy, and inference speed, addressing the trade-offs present in prior approaches. Additionally, human evaluations reaffirm its effectiveness in generating high-quality and contextually accurate audio content. These advancements position IMPACT as a robust solution for applications requiring both high fidelity and low latency.

\section*{Impact Statement}
This paper presents an approach to enhance text-to-audio generation. The key contributions include improving audio fidelity, quality, and inference speed through iterative mask-based parallel decoding applied to continuous latents within LDMs. 
Additionally, the method incorporates an unconditional pre-training phase, enabling the utilization of unlabeled audio data, which significantly enhances the fidelity, quality, and text relevancy of the generated audio.

\bibliography{example_paper}
\bibliographystyle{icml2025}

\newpage
\appendix
\onecolumn
\section{Classifier Free Guidance (cfg)}
\label{app:cfg}
\begin{figure*}[ht]
    \vskip 0.2in
    \centering
    \includegraphics[width=12cm]{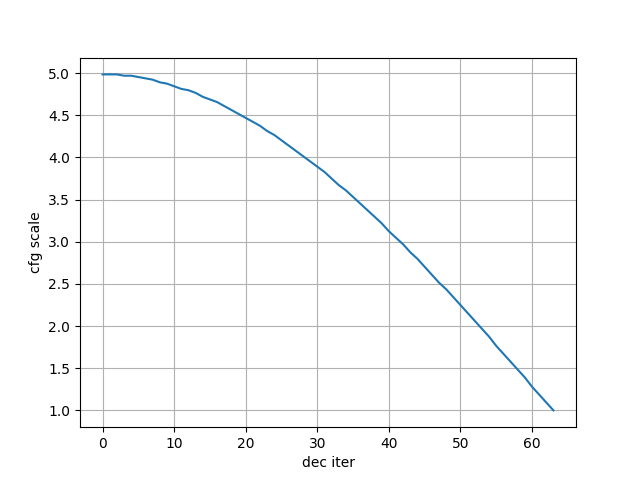}
    \caption{The cfg scaler $\beta_\text{cfg}^{(t)}$ at each decoding iteration during a 64-step iterative mask-based parallel decoding process.}
    \label{fig:cfg_sched}
    \vskip -0.1in
\end{figure*}

Cfg is achieved by pushing away the noise predictions $\epsilon_{\text{cond}}$ generated with the text condition from the ones $\epsilon_{\text{uncond}}$ generated without the text condition at each diffusion sampling step, as shown in Eq. (\ref{eq:cfg}).
\begin{equation}
    \label{eq:cfg}
    \epsilon = \epsilon_{\text{uncond}} + \beta_\text{cfg} \cdot \left( \epsilon_{\text{cond}} - \epsilon_{\text{uncond}} \right)
\end{equation}
The cfg scaler $\beta_\text{cfg}^{(t)}$ at decoding iteration $t$ controls how far to push away the generated latents from the unconditional ones. This scaler can be controlled by a cfg scheduler $\zeta^{(t)}$ during the iterative decoding process shown in Eq. (\ref{eq:cfg_sched}).
\begin{equation}
    \label{eq:cfg_sched}
    \beta_\text{cfg}^{(t)} = 1 + \zeta^{(t)} \cdot (\beta_\text{cfg\_max} - 1) 
\end{equation}
The cfg scheduler used in this work is $\zeta^{(t)} = \cos\left( \frac{\pi}{2} \cdot \frac{t}{T} \right)$.
The max cfg scaler $\beta_\text{cfg\_max}$ is set to 5.0 by default unless specified. The relationship between the cfg scaler $\beta_\text{cfg}^{(t)}$ and decoding iteration $t$ is visualized in Figure \ref{fig:cfg_sched}. Using a higher max cfg scaler indicates that the generated latents are more forced to follow the text condition during generation.

Table \ref{tab:cfg_1-5} reports the performance of using different max cfg scalers for IMPACT model (c) trained with AS (5500 h) for unconditional pre-training and AC+WC (1200 h) for text-conditional training. The results demonstrate that there is a high correlation between the max cfg scaler and the CLAP score.

\begin{table}[H]
    \caption{Performance of IMPACT with different max cfg scalers $\beta_\text{cfg\_max}$. }
    \label{tab:cfg_1-5}
    \vskip 0.15in
    \centering
    \begin{tabular}{ccccc}
        \toprule
        $\beta_\text{cfg\_max}$ & \textbf{FAD $\downarrow$} & \textbf{KL $\downarrow$} & \textbf{IS $\uparrow$} & \textbf{CLAP $\uparrow$} \\
        \midrule
        1.0 & 3.36 & 1.36 & 6.74  & 0.261 \\ 
        2.0 & 1.51 & 1.08 & 9.07  & 0.328 \\
        3.0 & 1.39 & \textbf{1.05} & 9.85  & 0.346 \\
        4.0 & 1.25 & 1.07 & \textbf{10.19} & \textbf{0.355} \\
        5.0 & \textbf{1.13} & 1.09 & 10.03 & 0.353 \\
        \bottomrule
    \end{tabular}
    \vskip -0.1in
\end{table}


\section{System Level Latency}
\label{app:latency}
\begin{table}[H]
    \caption{Latency values of baseline models and IMPACT base with different batch sizes measured in seconds on a single Tesla V100 GPU with 32 GB VRAM. ``diff steps'' indicates the number of diffusion sampling steps $\hat{t}$ used for diffusion-based models. ``coom'' is the abbreviation of cuda-out-of-memory, indicating that the GPU is unable to process the forwarding of the model with the corresponding batch size.}
    \label{tab:latency_all}
    \vskip 0.15in
    \centering
    \begin{tabular}{lcccccccc}
        \toprule
        \textbf{Batch Size} & diff steps $\hat{t}$ & \textbf{1} & \textbf{2} & \textbf{4} & \textbf{8} & \textbf{16} & \textbf{32} & \textbf{64} \\
        \midrule
        AudioGen v2 & - & 36.9 & 37.0 & 37.1 & 37.2 & 37.6 & 46.8 & 77.4 \\
        Tango 2 & 200 & 36.0 & 68.2 & 107.8 & 182.3 & coom & coom & coom \\
        EzAudio-L & 50 & 12.6 & 14.2 & 15.8 & 29.1 & 55.3 & 108.9 & coom\\
        EzAudio-XL & 50 & 14.4 & 14.5 & 21.4 & 40.2 & 78.4 & 155.4 & coom \\
        MAGNET-S & - & 1.6 & 2.1 & 3.7 & 6.9 & 13.3 & 26.2 & 52.0 \\
        MAGNET-L & - & 3.9 & 7.0 & 13.0 & 24.8 & 49.3 & 97.4 & 195.9 \\
        Make-an-Audio 2 & 100 & 3.5 & 11.1 & 12.6 & 15.9 & 34.1 & 41.3 & 81.7\\
        AudioLDM2-full & 200 & 46.7 & 48.8 & 57.8 & 96.1 & 148.2 & 275.4 & coom \\ 
        AudioLDM2-full-large & 200 & 77.9 & 78.2 & 153.7 & 195.7 & 328.1 & 643.7 & coom \\ \midrule
        IMPACT base, dec iter 64 & 100 & 19.8 & 19.9 & 20.6 & 22.2 & 24.0 & 27.9 & 37.3 \\
        IMPACT base, dec iter 32 & 100 & 10.1 & 10.4 & 10.5 & 11.2 & 12.6 & 15.3 & 20.5 \\
        IMPACT base, dec iter 16 & 100 & 5.0 & 5.2 & 5.4 & 5.7 & 6.6 & 8.5 & 13.1 \\
        IMPACT base, dec iter 8 & 100 & 2.5 & 2.6 & 2.8 & 3.2 & 4.1 & 5.7 & 9.5 \\
        IMPACT base, dec iter 4 & 100 & \textbf{1.3} & \textbf{1.4} & \textbf{1.5} & \textbf{1.8} & \textbf{2.7} & \textbf{4.4} & \textbf{7.8} \\
        
        \bottomrule
    \end{tabular}
    \vskip -0.2in
\end{table}

Table \ref{tab:latency_all} compares the latency values of baseline models and IMPACT base across various batch sizes (1, 2, 4, 8, 16, 32, 64) on a Tesla V100 GPU with 32 GB VRAM. Baseline models are configured using hyperparameters that yield their optimal performance on objective metrics. AudioGen v2\footnote{\href{https://github.com/facebookresearch/audiocraft/blob/main/model_cards/AUDIOGEN_MODEL_CARD.md}{https://github.com/facebookresearch/audiocraft/blob/main/model\_cards/AUDIOGEN\_MODEL\_CARD.md}} is the public version of AudioGen but is slightly different from the original AudioGen model published in \cite{kreuk2023audiogen}. AudioGen v2 generates audio in 10 seconds, adopts discrete representations from a retrained EnCodec model, and no audio mixing augmentations are used during training.

The results highlight the efficiency and scalability of IMPACT compared to existing baseline models. Notably, IMPACT demonstrates significantly lower latency across all batch sizes when using fewer decoding iterations. 

\section{Combinations of Decoding Iterations and Diffusion Steps of IMPACT base}
\label{app:dec_iter_diff_step}
\subsection{FAD}
\begin{figure*}[ht]
    \centering
    \includegraphics[width=10cm]{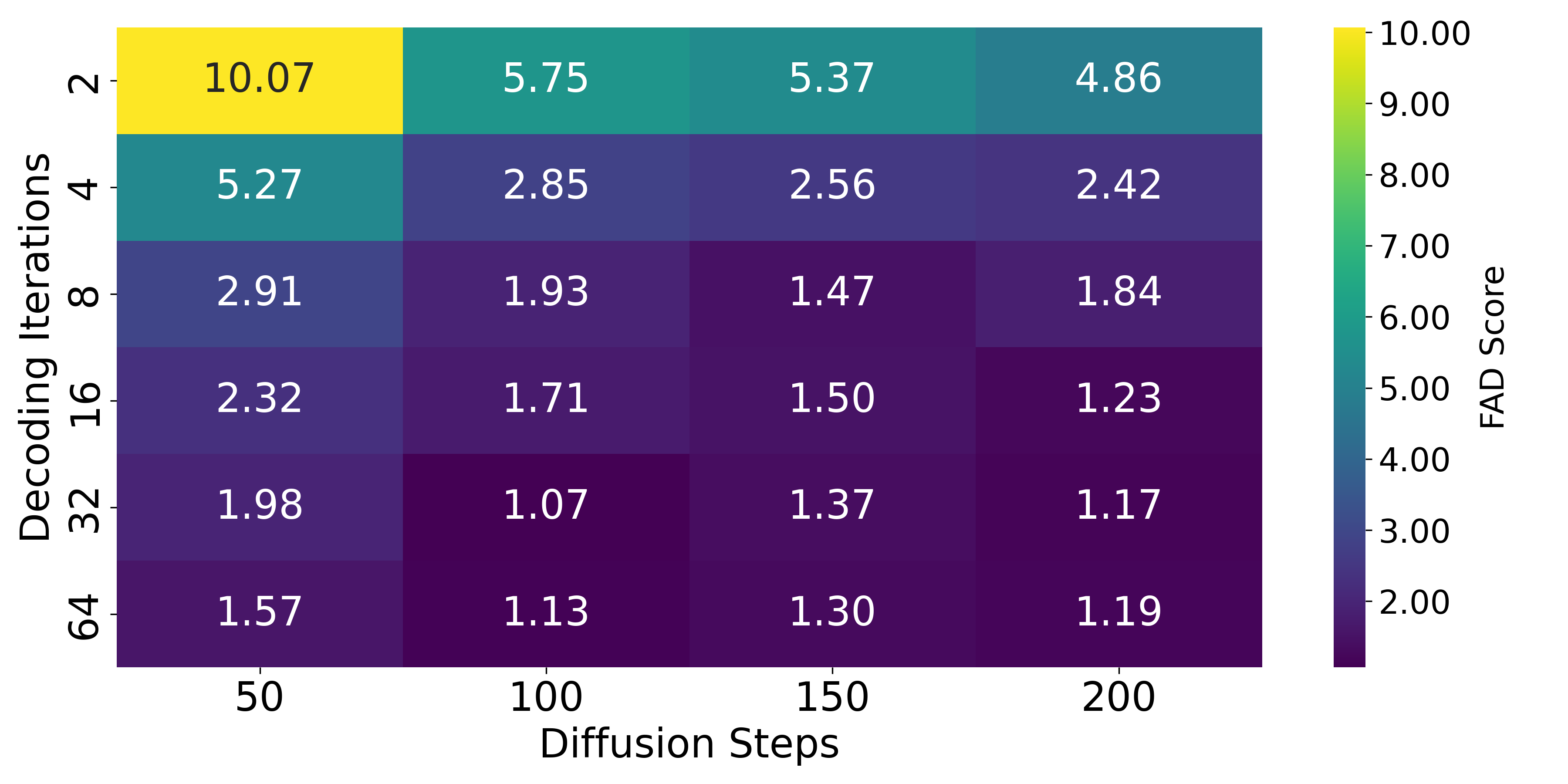}
    \caption{Heatmap visualizing FAD scores for IMPACT model (b) under varying decoding iterations and diffusion steps at a batch size of 8. FAD score is depicted by color intensity, with brighter colors indicating higher values.}
    \label{fig:dec_iter_diff_step_fad}
    \vskip -0.15in
\end{figure*}
\newpage
\subsection{KL}
\begin{figure*}[ht]
    \vskip 0.1in
    \centering
    \includegraphics[width=9cm]{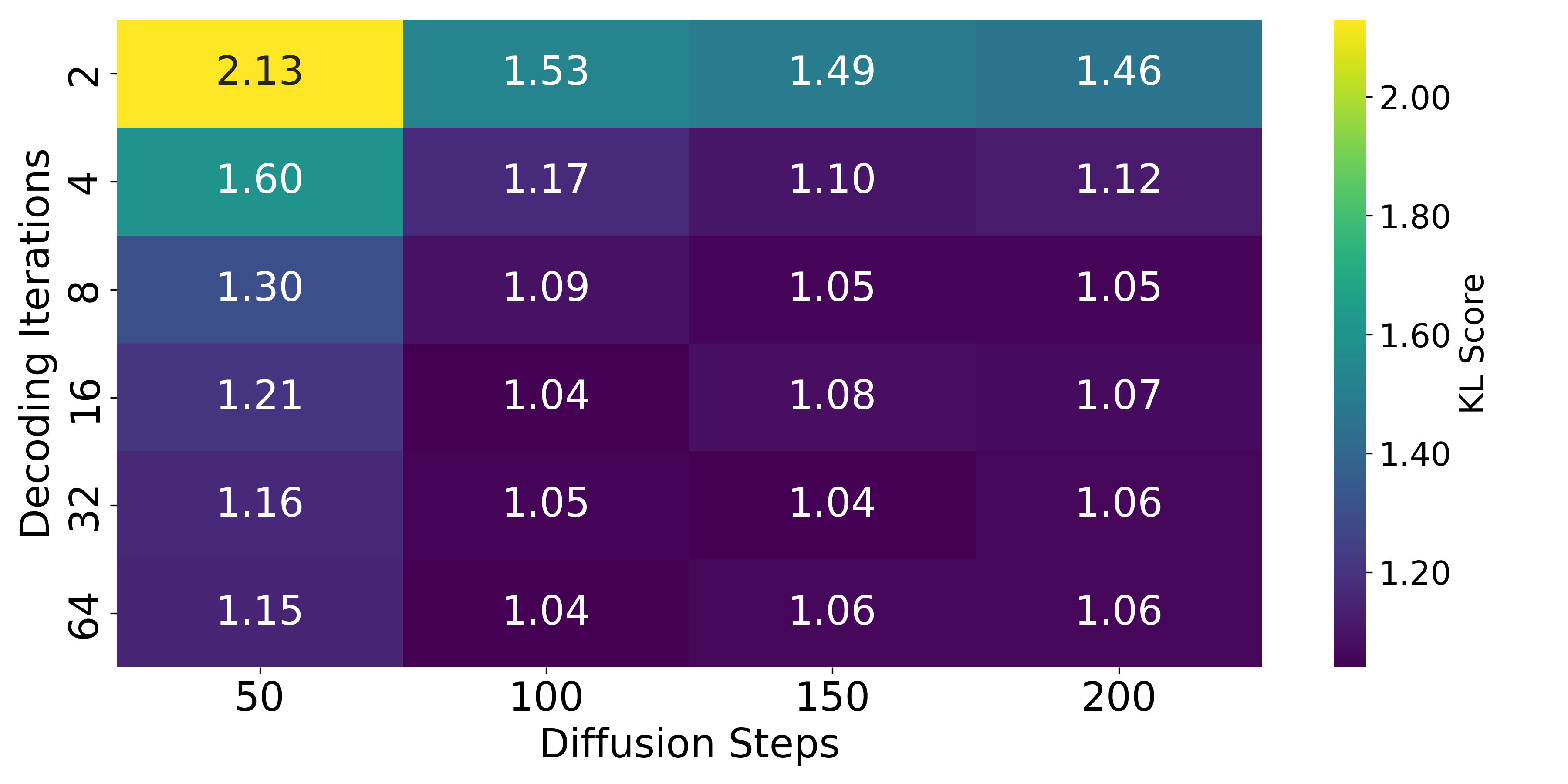}
    \vskip -0.1in
    \caption{Heatmap visualizing KL scores for IMPACT model (b) under varying decoding iterations and diffusion steps at a batch size of 8. KL score is depicted by color intensity, with brighter colors indicating higher values.}
    \label{fig:dec_iter_diff_step_kl}
    \vskip -0.15in
\end{figure*}

\subsection{IS}
\begin{figure*}[ht]
    \vskip 0.1in
    \centering
    \includegraphics[width=9cm]{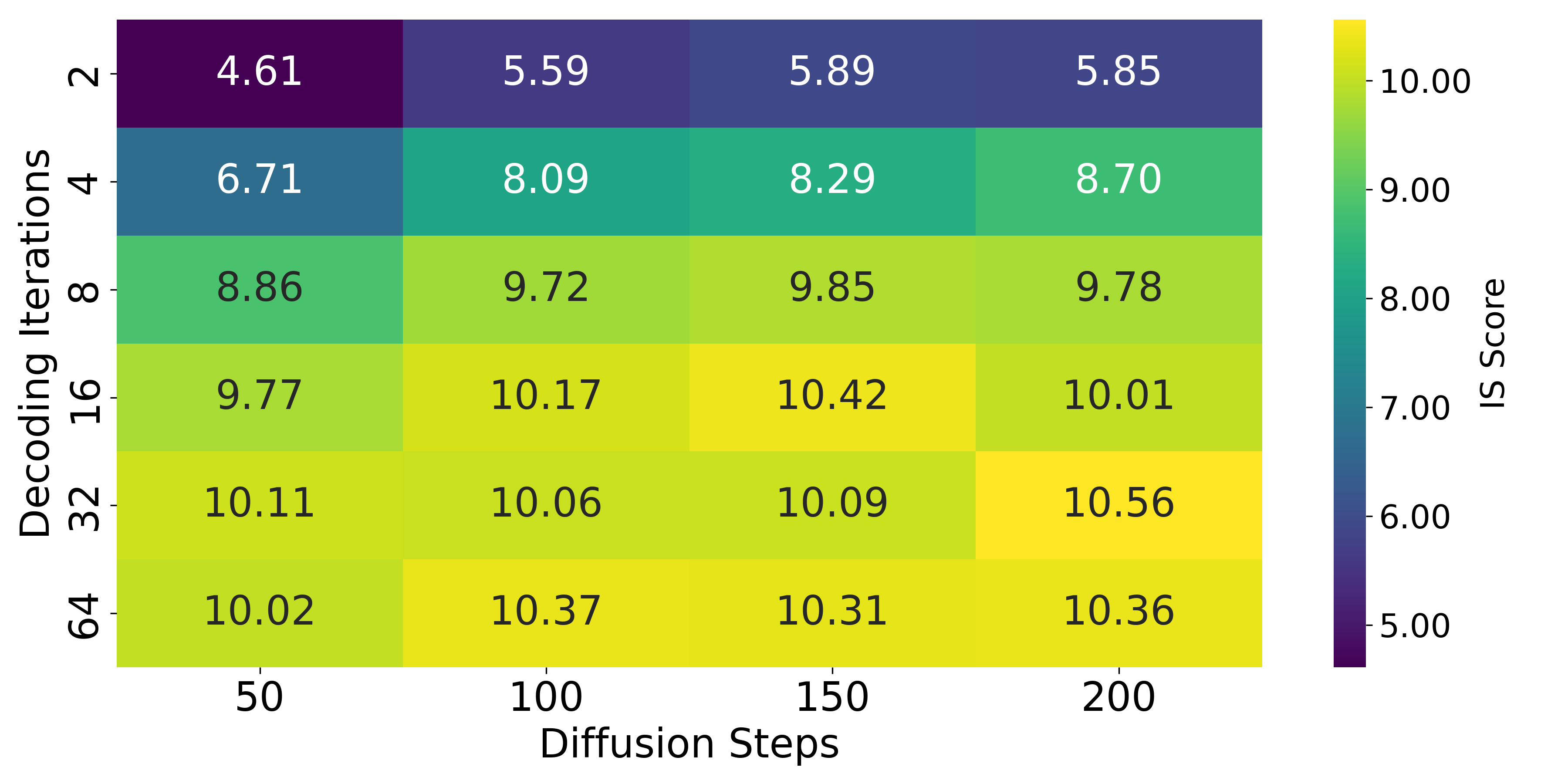}
    \vskip -0.1in
    \caption{Heatmap visualizing IS scores for IMPACT model (b) under varying decoding iterations and diffusion steps at a batch size of 8. IS score is depicted by color intensity, with brighter colors indicating higher values.}
    \label{fig:dec_iter_diff_step_is}
    \vskip -0.15in
\end{figure*}

\subsection{CLAP}
\begin{figure*}[ht]
    \centering
    \includegraphics[width=9cm]{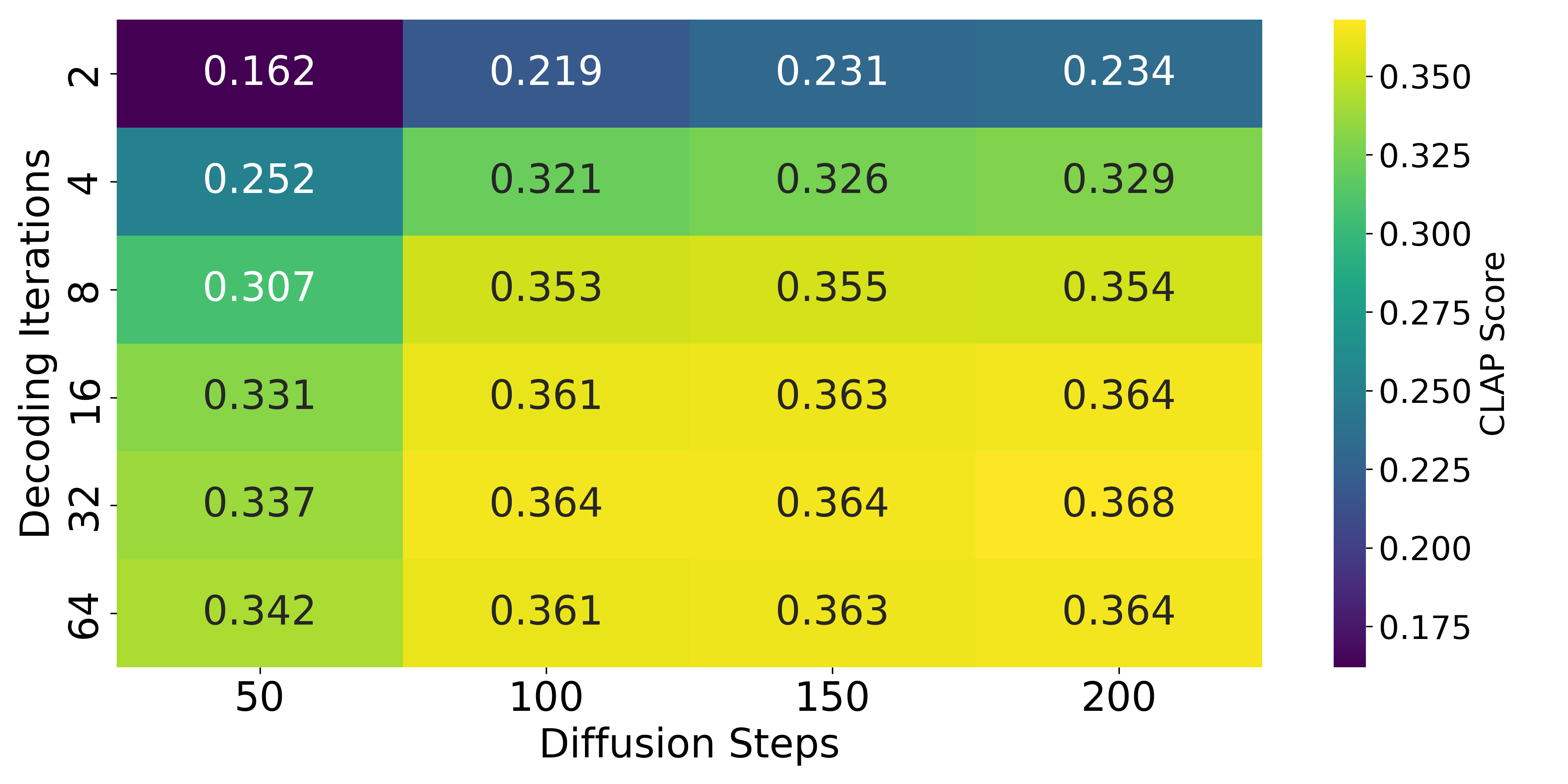}
    \vskip -0.1in
    \caption{Heatmap visualizing CLAP scores for IMPACT model (b) under varying decoding iterations and diffusion steps at a batch size of 8. CLAP score is depicted by color intensity, with brighter colors indicating higher values.}
    \label{fig:dec_iter_diff_step_clap}
    \vskip -0.15in
\end{figure*}


\newpage

\section{Implementation Details}
\label{app:modules}
\subsection{Random Position Selection}
\label{app:mask_stra}
Throughout the mask-based decoding process, the mask $M^{(t)}$ at each decoding iteration $t$ follows a subset property $M^{(t+1)} \subset M^{(t)}$ where $\subset$ indicates a strict subset in element-wise terms:
\begin{equation}
    \forall\ i \in \{1, \dots, N\}, \quad M^{(t+1)}[i] \leq M^{(t)}[i].
\end{equation}
This means a position $i$ that is unmasked ($M[i] = 0$) at decoding iteration $t$ must also have been unmasked at the rest of the decoding iterations. At each decoding iteration, the model predicts the positions that are different between $M^{(t+1)}$ and $M^{(t)}$, i.e., the positions where $M^{(t)}[i] = 1 \quad$ and $M^{(t+1)}[i] = 0$. We denote the set of indices of the positions to be predicted at each decoding iteration as $M^{(t)}_{\text{pred}}$. The behavior of this design at each decoding iteration $t+1$ will result in randomly selecting a subset of unpredicted positions of $\mathbf{z}^{(t)}$ to be generated.
This is different from previous mask-based generative models Soundstorm, VampNet, and MAGNET that operate on discrete audio tokens, since they are able to select positions that have low confidence scores to predict at each decoding iteration.

\subsection{Mask Scheduling}
\label{app:mask_sched}
Another purpose of applying the cosine function is that it is a concave function within $[0, \frac{\pi}{2}]$. This causes the number of positions to be predicted at decoding iteration $t$, denoted as $|M^{(t)}_{\text{pred}}|$, increasing throughout the iterative decoding process. This ensures that more latents are generated when more previous content is available.

\subsection{Transformer-based Latent Encoder}
\label{app:trans_enc}
The Transformer-based latent encoder $\text{Enc}_{\phi}$ consists of two subencoders, $\text{Enc}_{\phi_1}$ and $\text{Enc}_{\phi_2}$, which follows the scheme of a masked autoencoder (MAE) \cite{mae}\footnote{The original MAE paper treats this as an encoder-decoder framework. However, in this work, we refer to both the encoder and the decoder as encoders and name the whole module as ``transformer-based latent encoder'' since the role of this module is to encode the audio latents into a hidden representation $h$ to serve as the condition for the diffusion head.}. As shown in Eq. (\ref{eq:app_enc1}), the masked positions of the audio latents $\mathbf{z}$ are dropped before forwarding it to the first encoder $\text{Enc}_{\phi_1}$ along with the concatenated text condition vector sequence $\mathbf{c}$. We denote the audio latent sequence with masked positions dropped as $\mathbf{z}_{\text{drop}}$.
The reason for dropping the masked audio latents at this stage is to reduce both computational and memory requirements since a large portion of positions in $\mathbf{z}$ are masked.
\begin{equation}
    \label{eq:app_enc1}
    \mathbf{g} = \text{Enc}_{\phi_1}(\text{concat}(\mathbf{c}, \mathbf{z}_{\text{drop}} )), \mathbf{g} \in \mathds{R}^{(L + (1 - q) \cdot N) \times D}
\end{equation}
At this point, we reinsert mask latent embeddings (placeholder) to the masked positions in $\mathbf{g}$, resulting in $\mathbf{g}^\prime \in \mathds{R}^{(L + N) \times D}$.
$\mathbf{g}^\prime$ is then forwarded to $\text{Enc}_{\phi_2}$ to produce the final hidden representation $\mathbf{h}$ mentioned in Eq. (\ref{eq:h}). The forwarding process of $\text{Enc}_{\phi_2}$ is shown in Eq. (\ref{eq:app_enc2}).
\begin{equation}
    \label{eq:app_enc2}
    \mathbf{h} = \text{Enc}_{\phi_2}(\mathbf{g}^\prime), \mathbf{h} \in \mathds{R}^{(L + N) \times D}
\end{equation}

The Transformer blocks of the Transformer-based latent encoder are initialized using pre-trained MAR checkpoints\footnote{Checkpoints can be accessed from the official third-party MAR github repository \href{https://github.com/LTH14/mar?tab=readme-ov-file\#preparation.}{https://github.com/LTH14/mar?tab=readme-ov-file\#preparation}}, as failing to do so results in poor performance.

\subsection{Diffusion Head}
\label{app:diff_head}
In our work, we directly adopted the diffusion head of MAR's \cite{li2024autoregressive}. The diffusion head processes the input latents via a linear projection and then infuses time-dependent information through a timestep embedder. Condition vectors $h_i$ are likewise projected before being added to the time embedding, forming a combined representation that is passed through a series of ResBlock layers \cite{res}. Each ResBlock leverages adaptive layer normalization (AdaLN) \cite{peebles2023scalable}, where the conditioning vectors modulate the normalized hidden states via learned shifts and scales. Following these residual transformations, a final linear layer produces the output (e.g., mean and variance for diffusion). Unlike the Transformer-based latent encoder, this module is able to be trained from scratch.

\newpage

\section{Dataset Information}
\label{app:data}
Table \ref{tab:data_info} lists the pre-training data and text-conditional training data for each model. The pre-training data for IMPACT refers to the training data used for unconditional pre-training.

AS:AudioSet \cite{45857} , AC:AudioCaps \cite{kim-etal-2019-audiocaps}, WC:WavCaps \cite{10572302}, BBC:BBC sound effects, Cv2:Clotho v2 \cite{drossos2020clotho}, VGG:VGG-Sound, FSD50K:Freesound Dataset 50k\footnote{\href{https://annotator.freesound.org/fsd}{https://annotator.freesound.org/fsd}}, FS:Freesound Dataset, FTUS:Free To Use Sounds, SGE:Sonniss Game Effects, WSE:WeSoundEffects, PM:Paramount Motion, US:Urban Sound \cite{salamon2014dataset}, MI:Musical Instrument, MC:MusicCaps, GMG:Gtzan Music Genre, ESC50:Environmental Sound Classification \cite{piczak2015esc}, AA:Audio-alpaca\footnote{\href{https://huggingface.co/datasets/declare-lab/audio-alpaca}{https://huggingface.co/datasets/declare-lab/audio-alpaca}}, AFAS:AF-AudioSet, AACD:Auto-ACD \cite{autoacd}, ASQC:AS-Qwen-Caps, ASSLGC:AS-SL-GPT4-Caps, AASE:Adobe Audition Sound Effects\footnote{\href{https://www.adobe.com/products/audition/offers/adobeauditiondlcsfx.html}{https://www.adobe.com/products/audition/offers/adobeauditiondlcsfx.html}}, ASTK:Audiostock\footnote{\href{https://audiostock.net/}{https://audiostock.net/}}, MACS \cite{martin2021ground}, ES:Epidemic Sound\footnote{\href{https://www.epidemicsound.com/}{https://www.epidemicsound.com/}}, WT:WavText5Ks \cite{wavtext5k}, TUT:TUT acoustic scene \cite{mesaros2016tut}, FMA:Free Music Archive \cite{fma}, MSD: Million Song Dataset \cite{msd}, LJS:LJSpeech\footnote{\href{https://keithito.com/LJ-Speech-Dataset/}{https://keithito.com/LJ-Speech-Dataset/}}, GGS:GigaSpeech \cite{gigaspeech}.

\begin{table}[H]
\small
\scriptsize
\setlength\tabcolsep{3.0 pt}
\caption{Training data configurations for each model.}
\label{tab:data_info}
\vskip 0.15in
\centering
\begin{tabular}{llcc}
\toprule
& & \textbf{pre-train data} & \textbf{fine-tune data} \\
\midrule
&AudioGen & - & AS+BBC+AC+Cv2+VGG+FSD50K+FTUS+SGE+WSE+PM \\

&Tango & - & AC \\
&Tango-full-ft & AS+AC+FS+BBC+US+MI+MC+GMG+ESC50 & AC\\
&Tango-AF\&AC-FT-AC & AFAS, AC & AC \\
&Tango 2 & AS+AC+FS+BBC+US+MI+MC+GMG+ESC50 & AA \\
&EzAudio-L (24kHz) & AS+AACD+ASQC+ASSLGC & AC \\
&EzAudio-XL (24kHz) & AS+AACD+ASQC+ASSLGC & AC \\
&MAGNET-S & - & AS+BBC+AC+Cv2+VGG+FSD50K+FTUS+SGE+WSE+PM \\
&MAGNET-L & - & AS+BBC+AC+Cv2+VGG+FSD50K+FTUS+SGE+WSE+PM \\
&Make-an-Audio 2 & - & AS+AC+WC+AASE+ASTK+ESC50+FSD50K+MACS+ES+US+WT+TUT \\
&AudioLDM2-AC-large$^\star$ & - & AC \\
&AudioLDM2-full & - & AS+AC+WC+VGG+FMA+MSD+LJS+GGS \\
\midrule
\midrule
(a)& IMPACT base & - & AC+WC \\
(b)& IMPACT base  & AS & AC  \\
(c)& IMPACT base & AS  & AC+WC \\
(c')& IMPACT base & AS  & AC+WC \& AC \\
(d)& IMPACT base & AC+WC & AC+WC  \\ 
(d')& IMPACT base & AC+WC & AC+WC \\ \midrule
(e)& IMPACT large & AS & AC \\
(f)& IMPACT large & AS  & AC+WC \\

\bottomrule
\end{tabular}
\vskip -0.1in
\end{table}

\newpage
\section{Objective Evaluation}
\label{app:objective_metrics}
\subsection{FD and FAD}
FD and FAD are metrics specifically designed to assess the fidelity of generated audio by measuring the distance between the distributions of embeddings from real and generated audio samples. 
A lower FD or FAD score indicates that the generated audio closely resembles real audio in terms of these perceptual features, reflecting higher fidelity and realism. 
\subsection{KL}
KL divergence quantifies how the probability distribution of sound events in the generated audio differs from that of the real audio, with smaller values indicating that the generative model effectively captures the underlying distribution of the real audio data.
\subsection{IS}
IS measures both the quality and diversity of generated audio samples by computing the KL divergence between the conditional class distribution and the marginal class distribution over all samples using a pre-trained classifier. A higher IS suggests that the generated audio is high-quality and diverse. 

For metrics FD, FAD, KL, and IS, we follow the implementation of the commonly used audioldm\_eval\footnote{\href{https://github.com/haoheliu/audioldm_eval}{https://github.com/haoheliu/audioldm\_eval}} package.

\subsection{CLAP}
CLAP evaluates the semantic consistency between the input text and the generated audio by computing the cosine similarity between embeddings of input text prompts and generated audio in a shared embedding space learned by models trained to align textual descriptions with corresponding audio representations. A higher CLAP score signifies better alignment and that the generated audio accurately reflects the intended textual content. The CLAP model used for evaluation is clap-htsat-fused\footnote{\href{https://huggingface.co/laion/clap-htsat-fused}{https://huggingface.co/laion/clap-htsat-fused}}, which is different from the one\footnote{\href{https://huggingface.co/lukewys/laion_clap/blob/main/630k-audioset-fusion-best.pt}{https://huggingface.co/lukewys/laion\_clap/blob/main/630k-audioset-fusion-best.pt}} used for the text condition to avoid gaining advantage on the CLAP metric.

\newpage
\section{Subjective Evaluation Platform}
\label{app:human_eval}
\begin{figure*}[ht]
    \vskip 0.2in
    \centering
    \includegraphics[width=17cm]{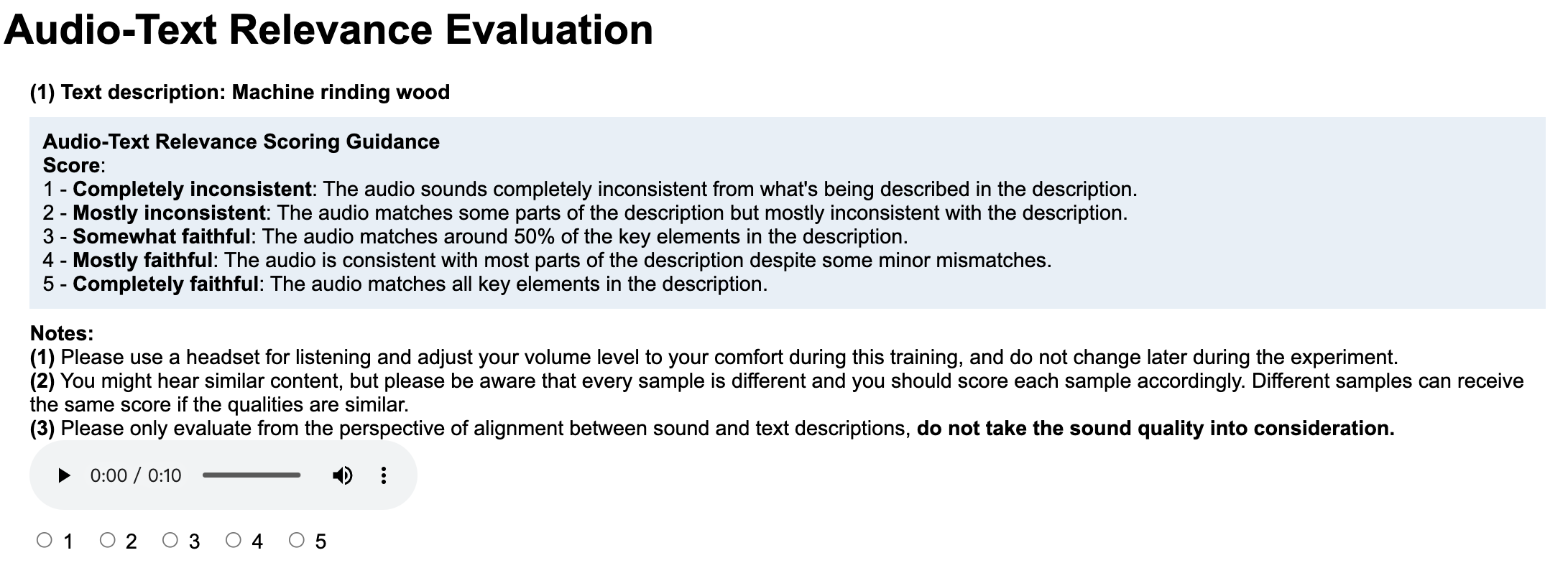}
    \caption{REL rating platform.}
    \label{fig:rel}
    \vskip -0.1in
\end{figure*}
\begin{figure*}[ht]
    \vskip 0.2in
    \centering
    \includegraphics[width=17cm]{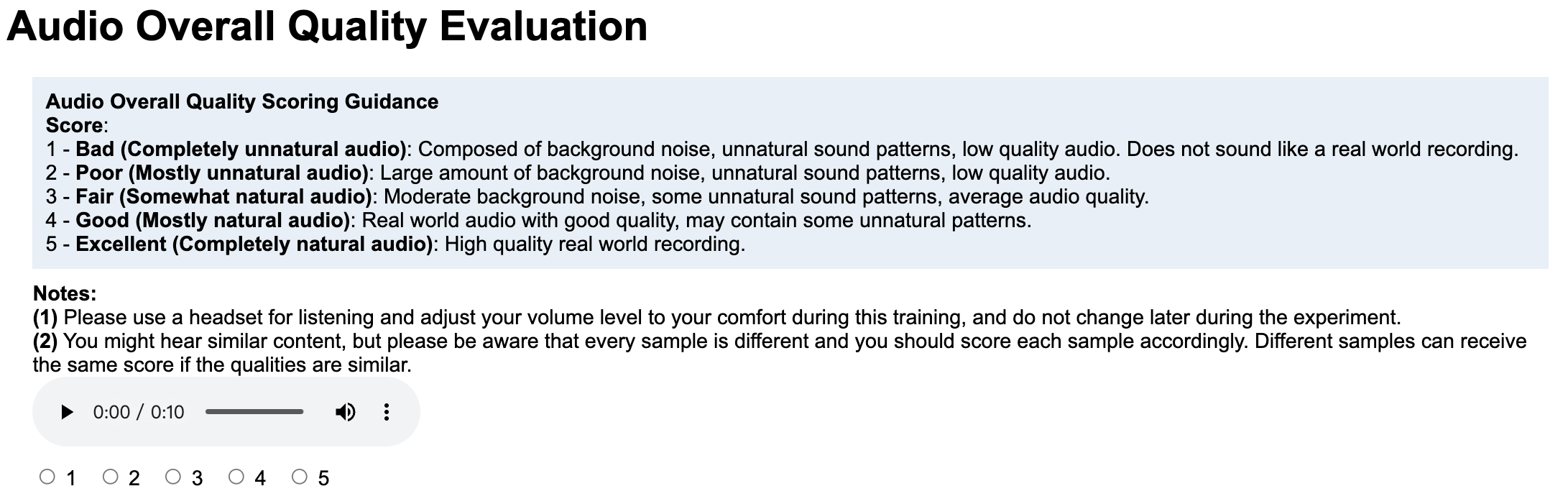}
    \caption{OVL rating platform.}
    \label{fig:ovl}
    \vskip -0.1in
\end{figure*}

\end{document}